\documentclass[twocolumn,prb,showpacs,multicol,amsmath,amssymb]{revtex4}
\usepackage[dvips]{graphicx}
\usepackage{graphicx}
\usepackage{dcolumn}
\usepackage{bm}
\usepackage{graphics}
\usepackage{epsfig,color}

\newcommand{\be}{\begin{equation}}
\newcommand{\ee}{\end{equation}}
\newcommand{\bea}{\begin{eqnarray}}
\newcommand{\eea}{\end{eqnarray}}

\begin{document}

\title{Numerical study of the one-dimensional quantum compass model}
\author{Saeed Mahdavifar}
\affiliation{ Department of Physics, University of
Guilan,41335-1914, Rasht, Iran }
\date{\today}

\begin{abstract}
The ground state magnetic phase diagram of the one-dimensional
quantum compass model (QCM) is studied using the numerical Lanczos
method. A detailed numerical analysis of the low energy excitation
spectrum is presented. The energy gap and the spin-spin
correlation functions are calculated for finite chains. Two kind
of the magnetic long-range orders, the  N$\acute{e}$el and a type
of the stripe-antiferromagnet, in the ground state phase diagram
are identified. Based on the numerical analysis, the first and
second order quantum phase transitions in the ground state phase
diagram are identified.
\end{abstract}


\maketitle

\section{Introduction}\label{sec1}

A theoretical understanding of the magnetic properties of
low-dimensional quantum magnets has attracted a lot of interest in
the last decades.  Especially, ground state properties are
interesting since the quantum fluctuation often plays the dominant
role at zero temperature. Recently, theoretical works are focused
on the important role played by the orbital degree of freedom in
determining the magnetic properties of materials. The complex
interplay among the spin and orbital degrees of freedom in quantum
magnets, such as Dzyaloshinskii-Moriya (DM)
interaction\cite{Dzyaloshinskii58,Moriya60}, makes their phase
diagram rich and induces various fascinating physical phenomena.

The experimental observations on Mott insulators are a realization
of the effect of the orbital degrees of freedom on the low-energy
behavior of a system. A good candidate for explaining the
low-temperature behavior of some Mott insulators is the quantum
compass model (QCM)\cite{Kugel73}.  In this model, the orbital
degrees of freedom are represented by (pseudo)spin-1/2 operators
and coupled anisotropically in such a way to mimic the competition
between orbital ordering in different directions. The
two-dimensional QCM is introduced as a realistic model to generate
protected cubits\cite{Doucot05} and can play a role in the quantum
information theory. This model is dual to studied model of
superconducting arrays\cite{Nussinov05}. It was shown that the
eigenstates are two-fold degenerate and to be
gapped\cite{Doucot05}. The results from both spin wave study and
exact diagonalization have suggested a first-order quantum phase
transition in the ground state phase diagram\cite{Dorier05}.
Recently, the existing of the first-order phase transition is
confirmed in two-dimensional QCM\cite{Chen07,Otus09}. The dilution
effect is studied by means of the quantum Monte-Carlo
method\cite{Tanaka07}. It is found that due to dilution, the
ordering temperature decreases much stronger than that in spin
models. A numerical monte carlo simulation also is done on the
square lattice and critical temperatures are obtained for the
classical and quantum versions\cite{Wenzel08}.
\begin{figure}[t]
\centerline{\psfig{file=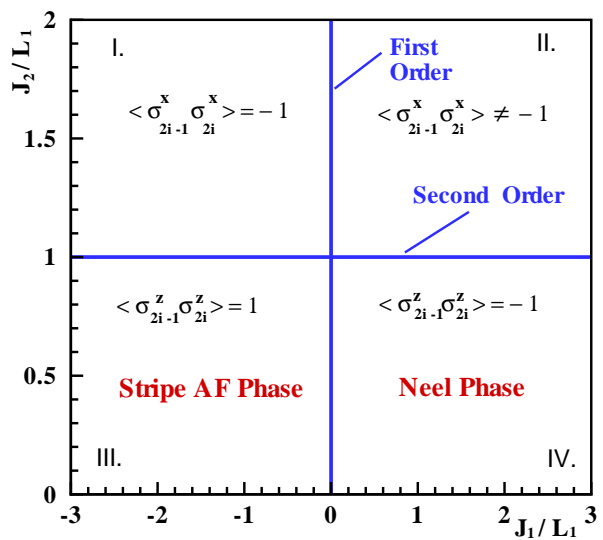,width=3.65in}}
\caption{(Color online.) The phase diagram of the one-dimensional
quantum compass model. The first-order and the second-order
critical lines denote with $J_{1}/L_1=0$ and $J_{2}/L_1=1$,
respectively. The short-range correlation functions in different
regions are given by
$\langle\sigma_{2i-1}^{x(z)}\sigma_{2i}^{x(z)}\rangle$. In the
regions III. and IV., there are the stripe AF and N$\acute{e}$el
long-range orders, respectively.} \label{phase-diagram}
\end{figure}

The one-dimensional (1D) QCM has been studied much
less\cite{Brzezicki07,You08,Brzezicki09,Sun1-09,Sun2-09,Eriksson09}.
The Hamiltonian of the 1D QCM on a periodic chain of $N$ sites is
given by\cite{Eriksson09}
\begin{eqnarray}
{\cal H} = \sum_{i=1}^{N/2} J_{1} \sigma_{2i-1}^{z}\sigma_{2i}^{z}
+ J_{2}\sigma_{2i-1}^{x} \sigma_{2i}^{x}+ L_{1} \sigma_{2i}^{z}
\sigma_{2i+1}^{z}. \label{Hamiltonian}
\end{eqnarray}
Here $\sigma_{i}^{x,z}$ are the Pauli operators on the $i$th site
and $J_1, J_2, L_1$ are the exchange couplings. By mapping the
model to a quantum Ising model, an exact solution is obtained for
the ground state energy and the complete excitation
spectrum\cite{Brzezicki07}. It is shown that the 1D QCM exhibits a
first-order phase transition at $J_1=0$ between two disordered
phases with opposite signs of certain local spin correlations. The
degeneracy of the ground state energy at this transition point is
$2\times2^{N/2}$ in the limit of $N\rightarrow\infty$ and the
energy gap is closed. The model is also diagonalized exactly by a
direct Jordan-Wigner transformation\cite{Brzezicki09}. The
obtained results by latter approach, confirm the existence of the
first-order phase transition in the ground state phase diagram.
In a very recent paper, Eriksson et.al.,\cite{Eriksson09}  have
shown that the reported first-order phase transition, in fact
occurs at a multicritical point where a line of the first-order
transition ($J_1/L_1=0$) meets with a line of the
second-order($J_2/L_1=1$) transition. Also, there are four gapped
phases in the regions: (I.) $J_1/L_1<0$, $J_2/L_1>1$, (II.)
$J_1/L_1>0$, $J_2/L_1>1$, (III.) $J_1/L_1<0$, $J_2/L_1<1$, (IV.)
$J_1/L_1>0$, $J_2/L_1<1$ (see Fig.\ref{phase-diagram}). Since the
spin ordering in different sectors is obtained by study of
short-range correlation functions
($\langle\sigma_{2i-1}^{x,z}\sigma_{2i}^{x,z}\rangle$), the
complete picture of different long-range orders of the 1D QCM is
unclear up to now.

\begin{figure}
\centerline{\psfig{file=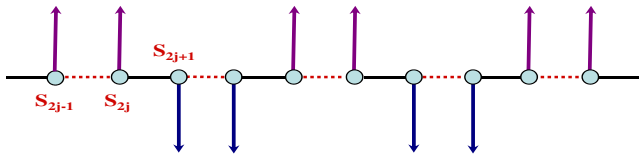,width=2.8in,height=2.0cm,
angle=0}} \caption{(Color online.) The Schematic picture of the
canted spins in the direction of the $z$ axis, which is known as
the saturated stripe-antiferromagnetic order. } \label{Schematic}
\end{figure}

For the first time, in this paper we perform an accurate numerical
experiment on the ground state phase diagram of the model. By
analyzing the numerical results on finite chains, we draw a clear
picture of the different long-range orders in the ground state
phase diagram. In particular we apply the Lanczos method to
diagonalize numerically finite ($N=12, 16, 20, 24$) chain systems.
Using the exact diagonalization results, we calculate the various
spin structure factors.
Based on the exact diagonalization results we argue that instead
of the analytical suggested N$\acute{e}$el order, the
stripe-antiferromagnetic long-range order (see
Fig.\ref{Schematic}) exist in the region (III.) $J_1/L_1<0$,
$J_2/L_1<1$ of the ground state phase diagram. We denote by
"stripe-antiferromagnetic phase" the phase with the opposite
magnetization along the $z$ axis on the odd bonds.

The paper is organized as follows. In the forthcoming section we
present numerical results of our exact diagonalization studies of
the system. In Section III, we conclude and summarize our results.

\section{Numerical results} \label{sec2}

In this section, to explore the nature of the spectrum and the
quantum phase transition, we used Lanczos method to diagonalize
numerically chains with length up to $N=24$ and different values
of the exchanges (a) $J_{2}/L_1=2(1-J_1/L_1)$, (b)
$J_{2}/L_1=1-J_1/L_1$ and (c) $J_{2}/L_1=(1/2)(1-J_1/L_1)$. The
energies of the few lowest eigenstates were obtained for chains
with periodic boundary conditions. The Lanczos method and the
related recursion methods\cite{Lanczos50,Grosso95}, possibly with
appropriate implementations, have emerged as one of the most
important computational procedures, mainly when a few extreme
eigenvalues are desired.

\begin{figure}
\centerline{\psfig{file=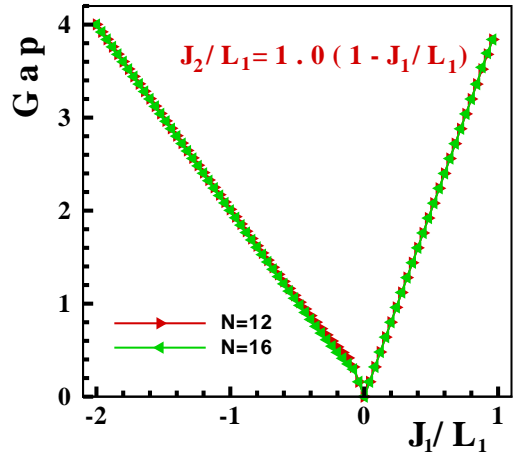,width=3.0in}}
\caption{(Color online.) Difference between the energy of the
first excited state and the ground state energy as a function of
the exchange $J_1$, for chain exchanges $L_{1}=1$,
$J_2/L_1=1-J_1/L_1$, including different lengths $N=12, 16$. }
\label{energy-gap}
\end{figure}
\begin{figure*}[t]
\centerline{\includegraphics[width=6cm,height=6cm,angle=0]{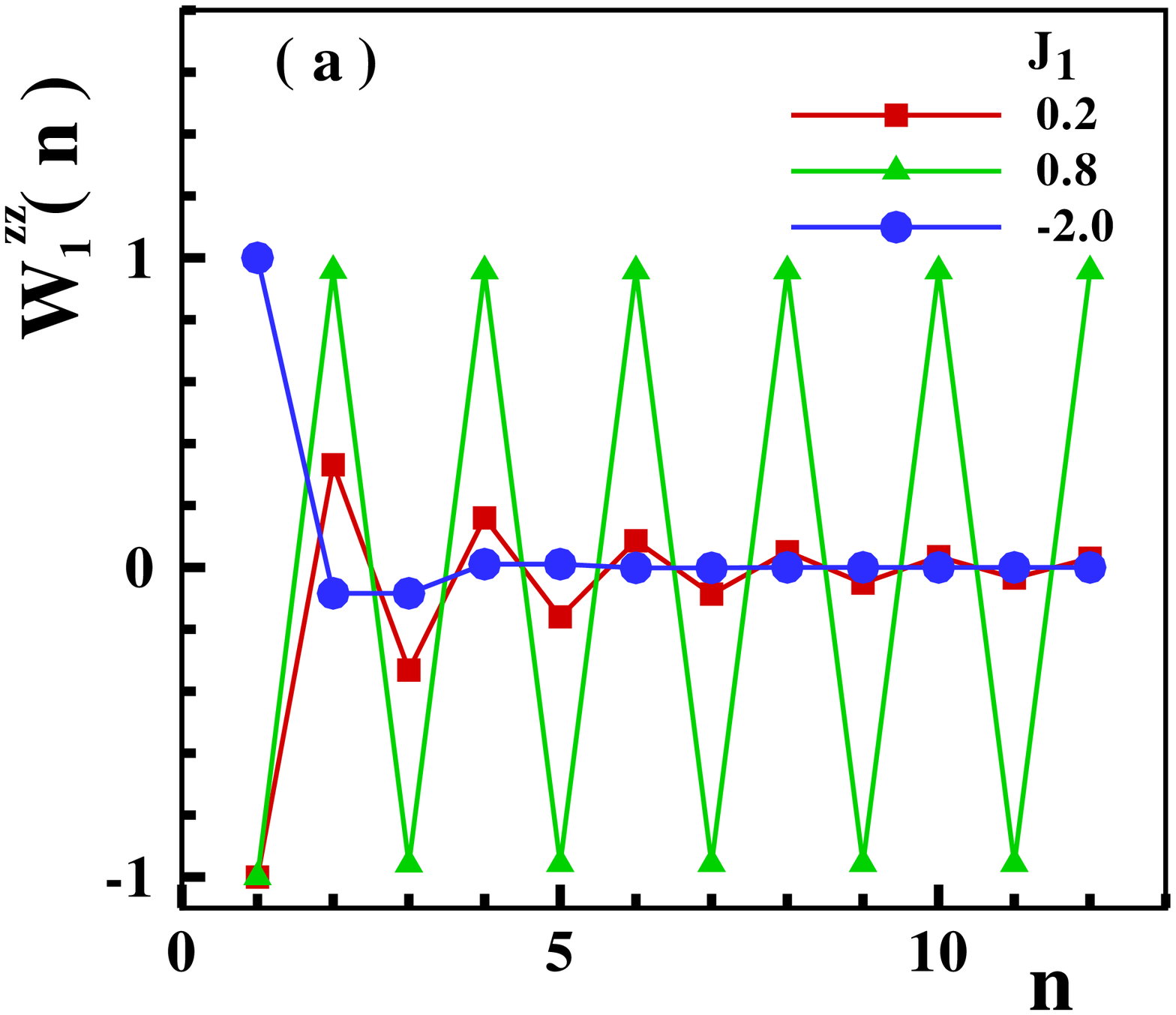}
\includegraphics[width=6cm,height=6cm,angle=0]{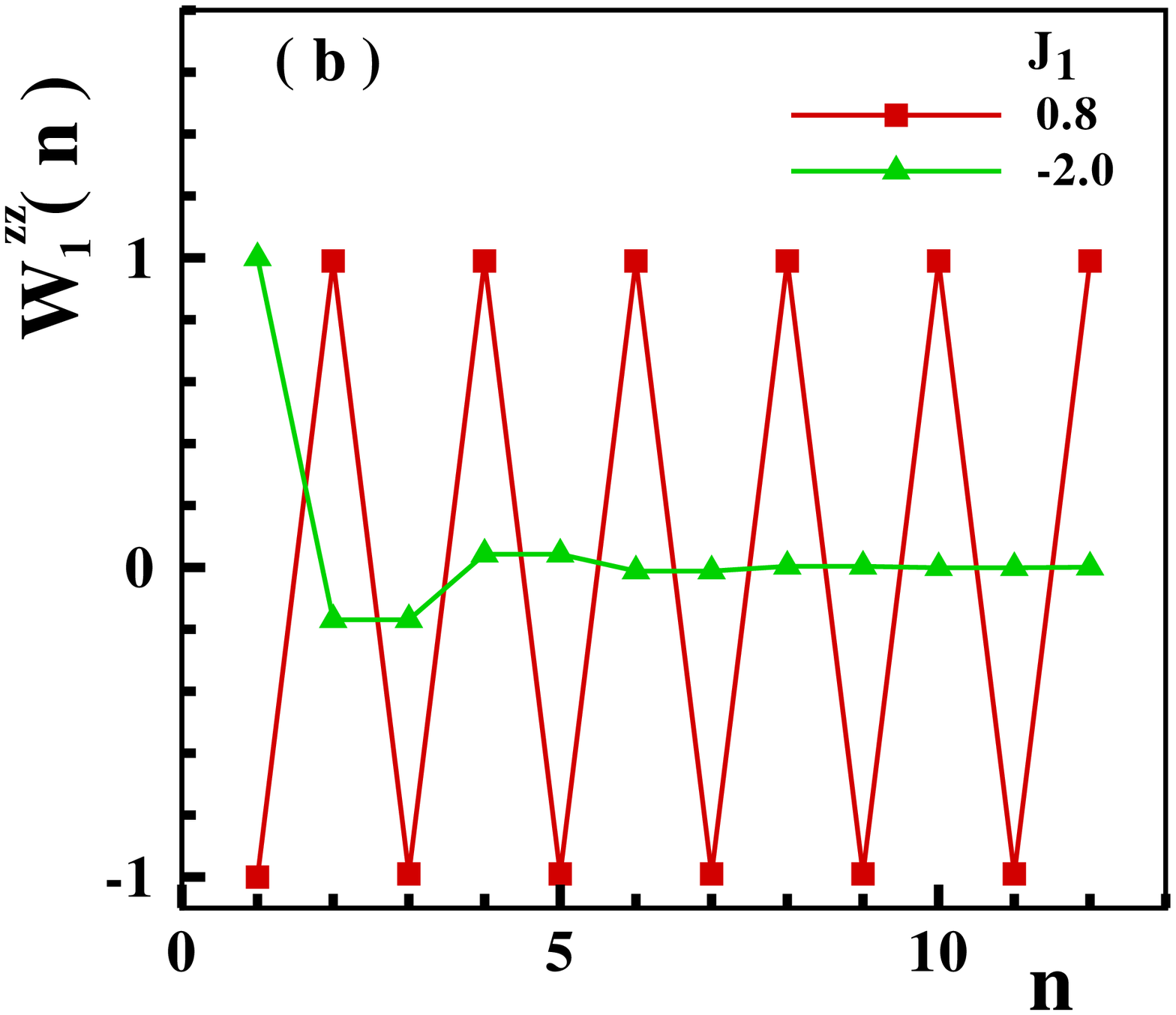}
\includegraphics[width=6cm,height=6cm,angle=0]{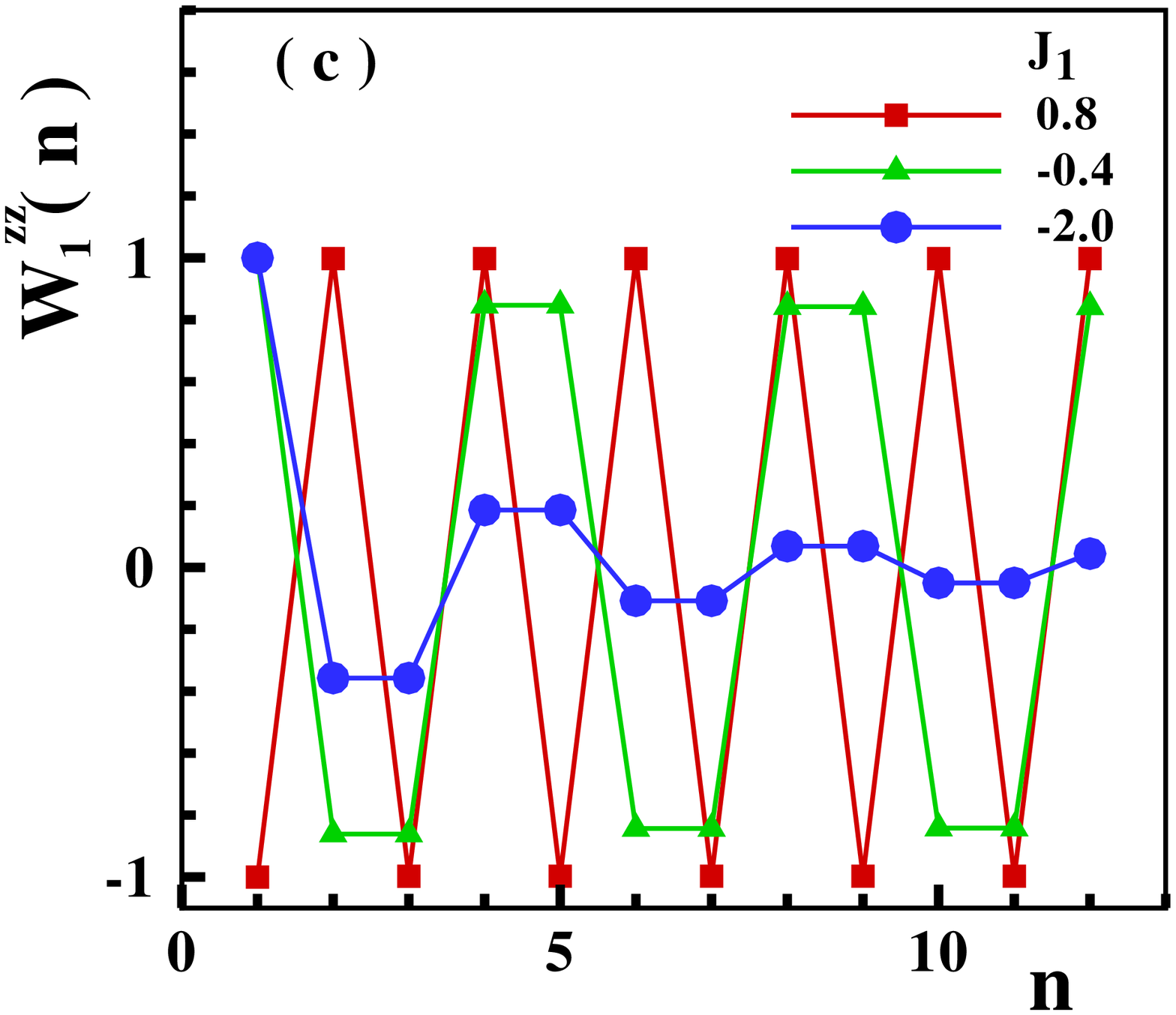}
} \caption{(Color online.) The spin correlation $W_{1}^{zz}(n)$ as
a function of the $n$, for chain length $N=24$ and exchanges
$L_{1}=1.0$, (a)$J_2/L_1=2(1-J_1/L_1)$, (b)$J_{2}/L_1=1-J_1/L_1$
and (c) $J_{2}/L_1=(1/2)(1-J_1/L_1)$. } \label{correlations}
\end{figure*}
\subsection{The Energy Gap}\label{subsec-2a}

To show that the transition lines could be observed from the
numerical calculations of small systems we start our consideration
with the energy gap function. First, we have computed the three
lowest energy eigenvalues of chains with  different values of the
exchanges $J_{2}/L_1=1-J_1/L_1$.   To get the energies of the few
lowest eigenstates we consider chains with periodic boundary
conditions. In Fig.\ref{energy-gap}, we present results of these
calculations for the exchanges $L_1=1$, $-2<J_1/L_1<1$ and chain
sizes $N=12, 16$. We define the excitation gap as a gap to the
first excited state. As it can be seen in Fig.\ref{energy-gap}, in
the considered limit of exchanges, this difference is
characterized by the indistinguishable (within the used numerical
accuracy) dependence on the chain length and shows an universal
linear decrease with increasing exchange $J_1/L_1$. At
$J_1/L_1=-2$ the spectrum model is gapped. By increasing the
exchange $J_1/L_1$ from $-2$, the energy gap decreases linearly
with $J_1/L_1$.
The energy gap vanishes at $J_1/L_1=0$, which is the only level
crossing between the ground-state energy and the first excited
state energy. With more increasing $J_1/L_1$, the spin gap opens
again and for a sufficiently large exchange $|J_1|/L_1$, it
increases linearly. Therefore, in complete agreement with the
analytical results\cite{Eriksson09}, we conclude that in this
case, there are two different gapped phases in the regions (I.)
$J_1/L_1<0$, $J_2/L_1>1$, and (IV.) $J_1/L_1>0$, $J_2/L_1<1$.


\subsection{The ground state phase diagram}\label{subsec-2b}

To recognize the different phases induced by the exchanges $J_1$
and $J_2$ in the ground state phase diagram of the 1D QCM, we
implemented the algorithm for finite-size chains ($N=12, 16, 20,
24$) to calculate the order parameters and the various spin
correlation functions.

Because of two type of links, we have computed two kind of
two-point (short-range) correlation functions defined as
$\langle\sigma_{2i-1}^{x(z)}\sigma_{2i}^{x(z)}\rangle$,
$\langle\sigma_{2i}^{x(z)}\sigma_{2i+1}^{x(z)}\rangle$. Where
$\langle...\rangle$ denotes the ground state average.
In complete agreement with the analytical
results\cite{Eriksson09}, found four regimes with different
short-range correlations. In the region $J_2/L_1<1$, antiparallel
ordering of spin $z$ component on even bonds
($\langle\sigma_{2i}^{z}\sigma_{2i+1}^{z}\rangle$) is dominated
with respect to the  antiparallel ordering of spin $x$ component
on odd bonds ($\langle\sigma_{2i-1}^{x}\sigma_{2i}^{x}\rangle$) in
the region $J_2/L_1>1$. On the other hand, in the region
$J_1/L_1<0$ the ground state is in the $s=N/2$ subspace with
$\langle\sigma_{2i-1}^{z}\sigma_{2i}^{z}\rangle=1$ and  in the
region $J_1/L_1>0$, is in the $s=0$ subspace with
$\langle\sigma_{2i-1}^{z}\sigma_{2i}^{z}\rangle=-1$. Therefore,
the first-order quantum phase transition at $J_1/L_1=0$
corresponds to the flip sign of the $z$ component of the pair
correlation functions of spins on every first bonds
($\langle\sigma_{2i-1}^{z}\sigma_{2i}^{z}\rangle$). In following
we try to draw a picture of the magnetic long-range ordering of
the system in different ground state phases.

To find the long-range magnetic order of the ground state of the
system, we start our consideration with the spin-spin correlation
function defined by
\begin{eqnarray}
W_{j}^{\alpha\alpha}(n) =
\langle\sigma_{j}^{\alpha}\sigma_{j+n}^{\alpha}\rangle~~~(\alpha=x,z),
\end{eqnarray}
and the spin structure factor at momentum $q$ defined by
\begin{eqnarray}
S^{\alpha\alpha}(q) = \sum_{n=1}^{N-1}W_{j}^{\alpha\alpha}(n)
exp(iqn).
\end{eqnarray}
\begin{figure*}[t]
\centerline{\includegraphics[width=6cm,height=6cm,angle=0]{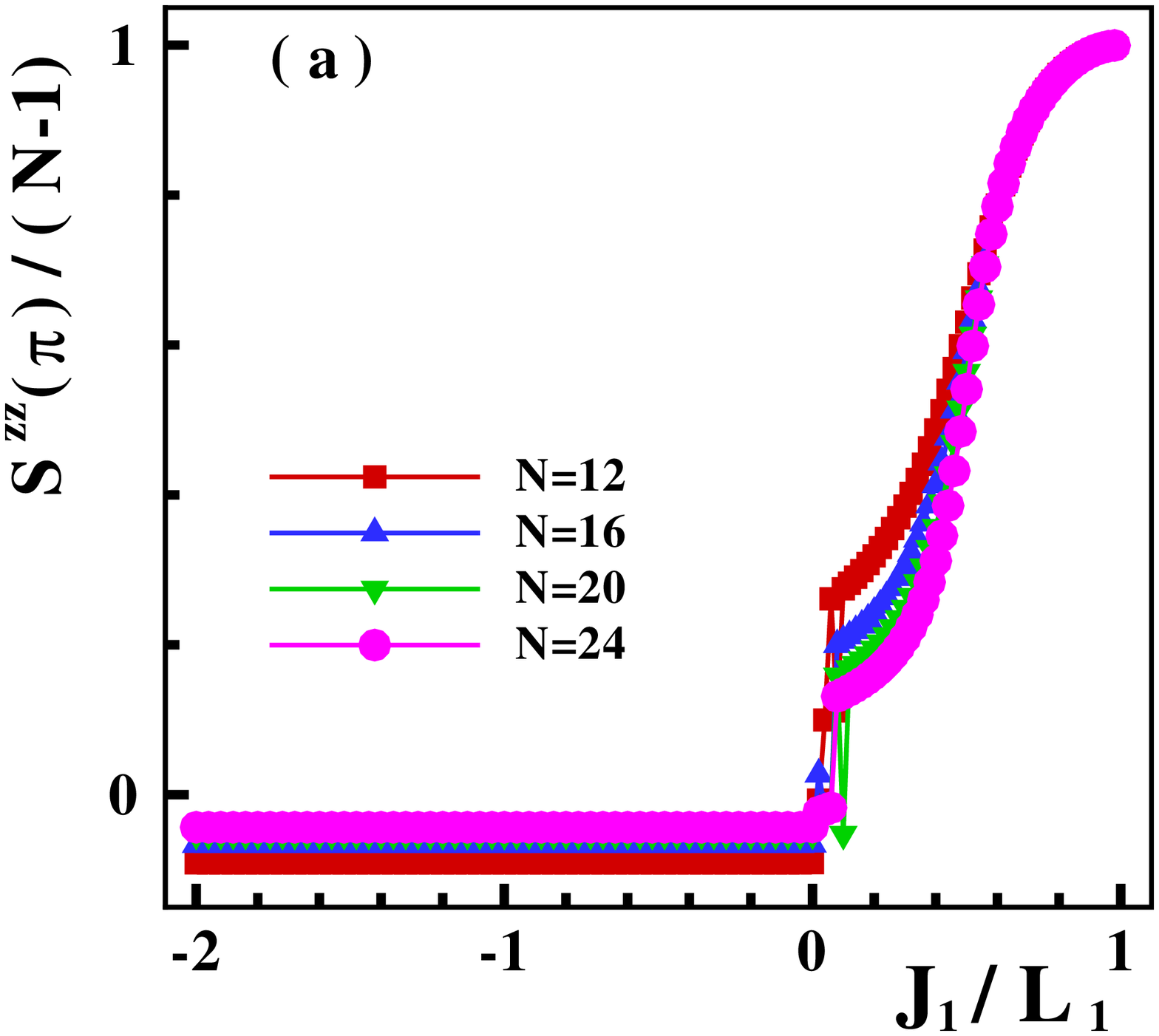}
\includegraphics[width=6cm,height=6cm,angle=0]{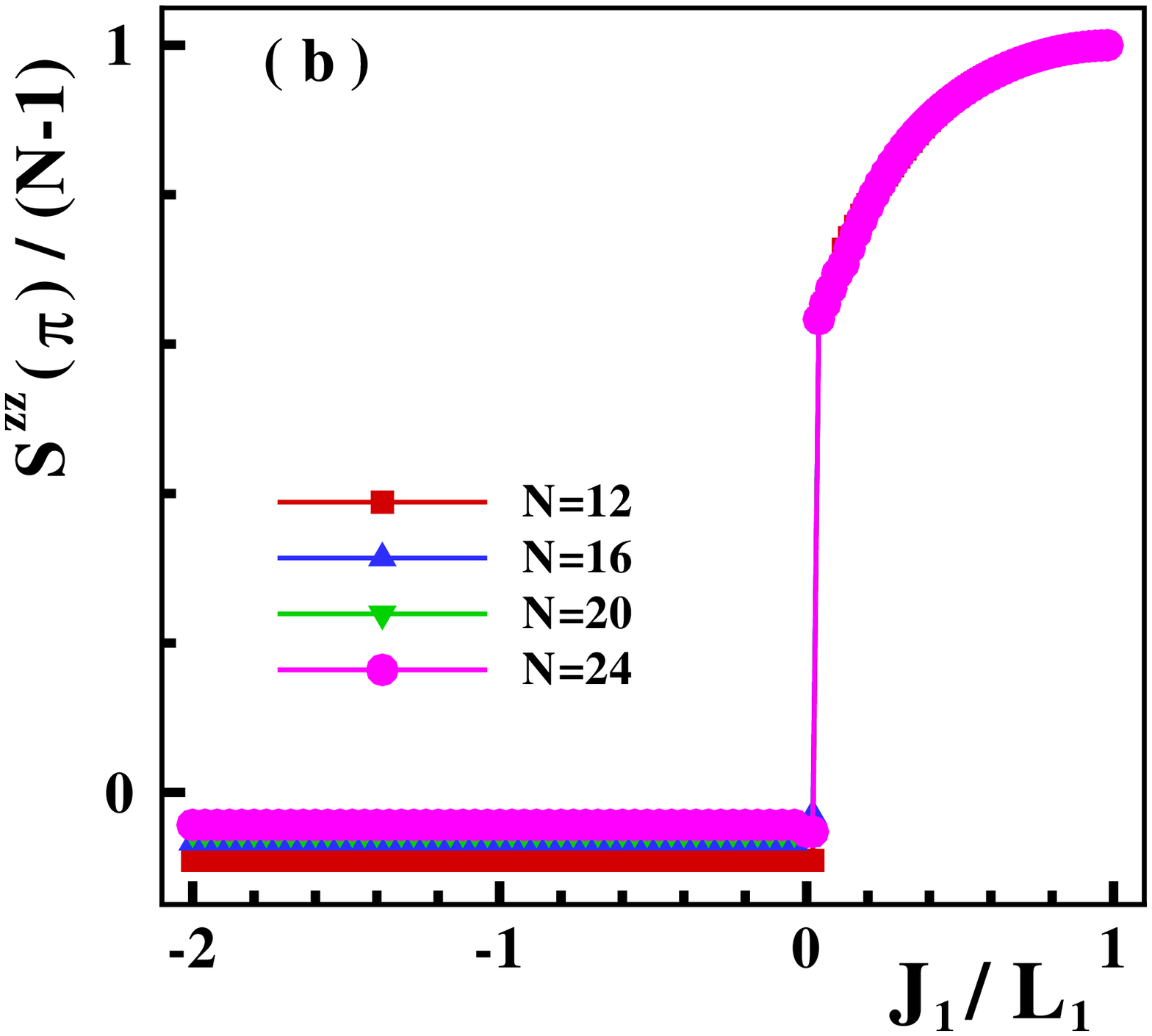}
\includegraphics[width=6cm,height=6cm,angle=0]{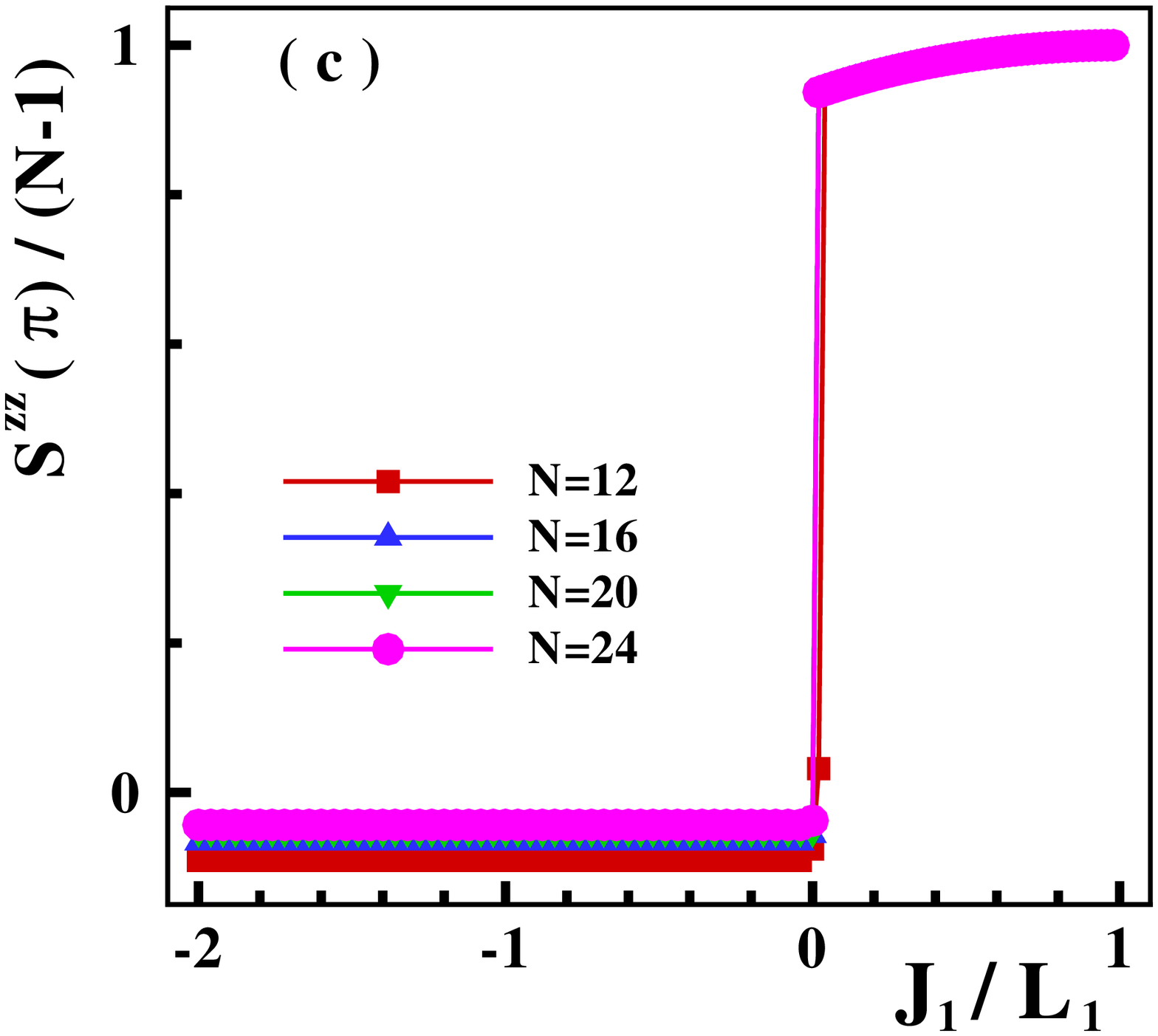}
} \caption{(Color online.) The spin structure factor $S^{zz}(\pi)$
as a function of $J_1$, for chain lengths $N=12, 16, 20, 24$ and
exchanges $L_{1}=1.0$, (a)$J_2/L_1=2(1-J_1/L_1)$,
(b)$J_{2}/L_1=1-J_1/L_1$ and (c) $J_{2}/L_1=(1/2)(1-J_1/L_1)$. }
\label{Str-Neel}
\end{figure*}
\begin{figure*}[t]
\centerline{\includegraphics[width=6cm,height=6cm,angle=0]{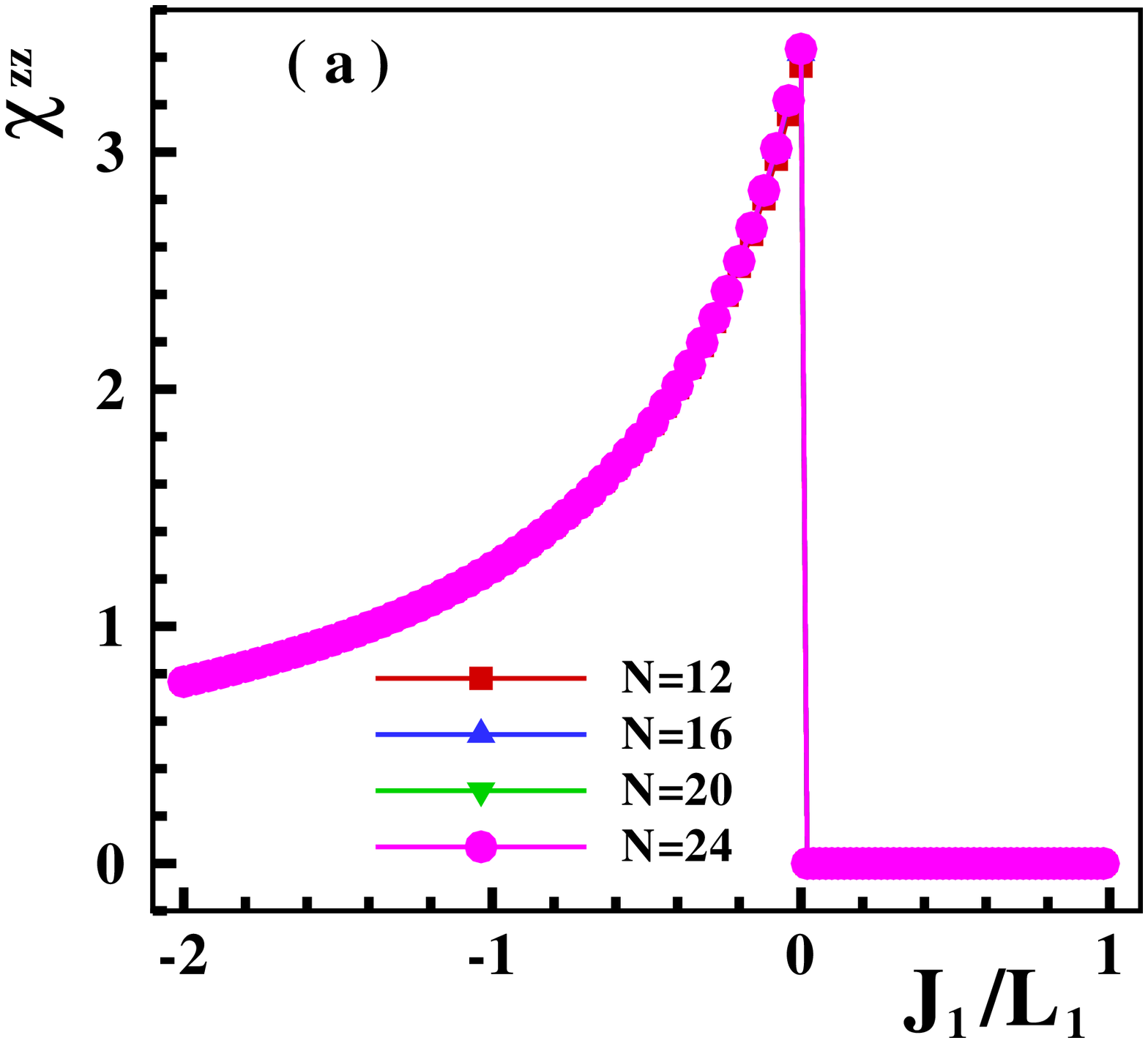}
\includegraphics[width=6cm,height=6cm,angle=0]{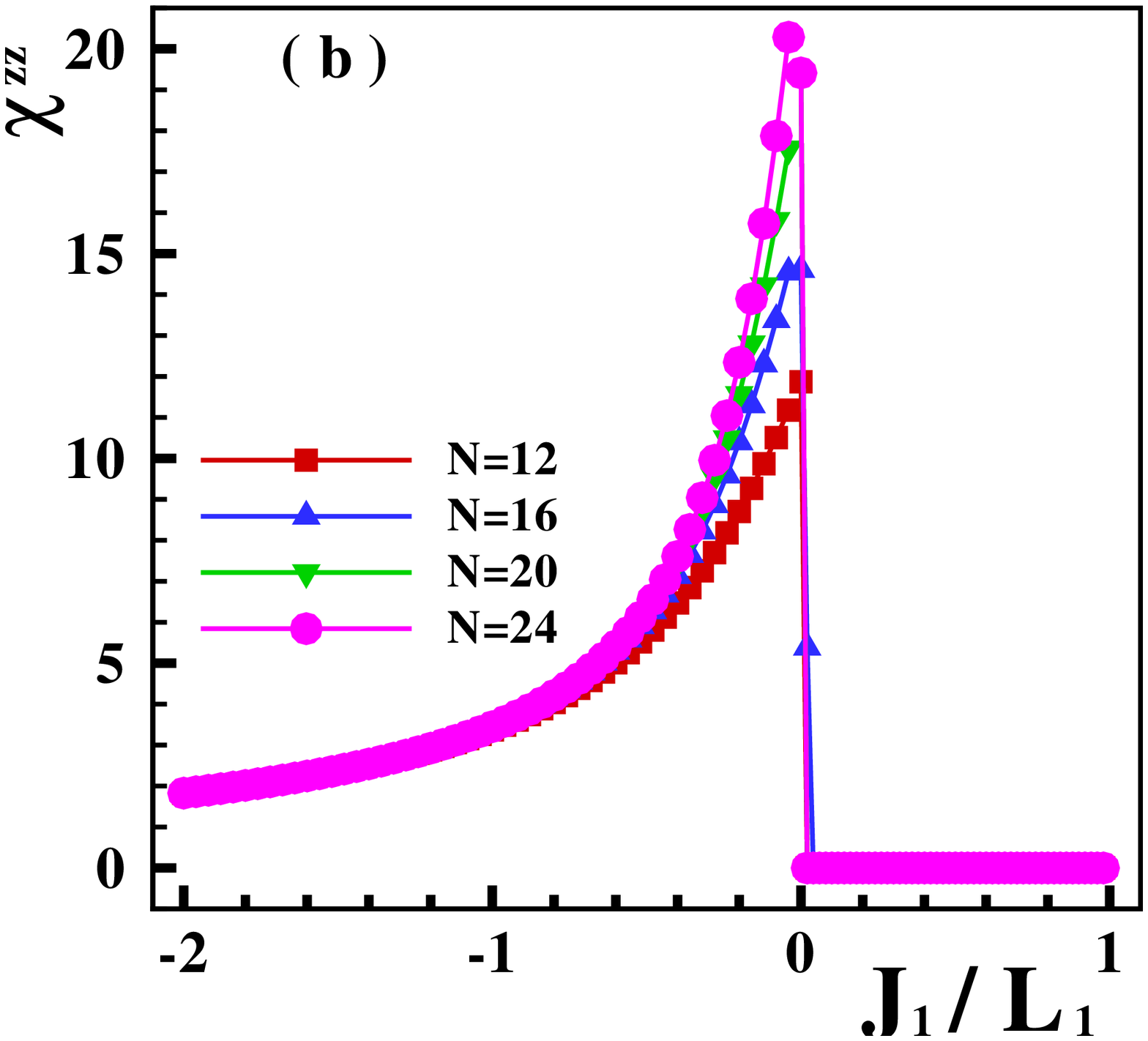}
\includegraphics[width=6cm,height=6cm,angle=0]{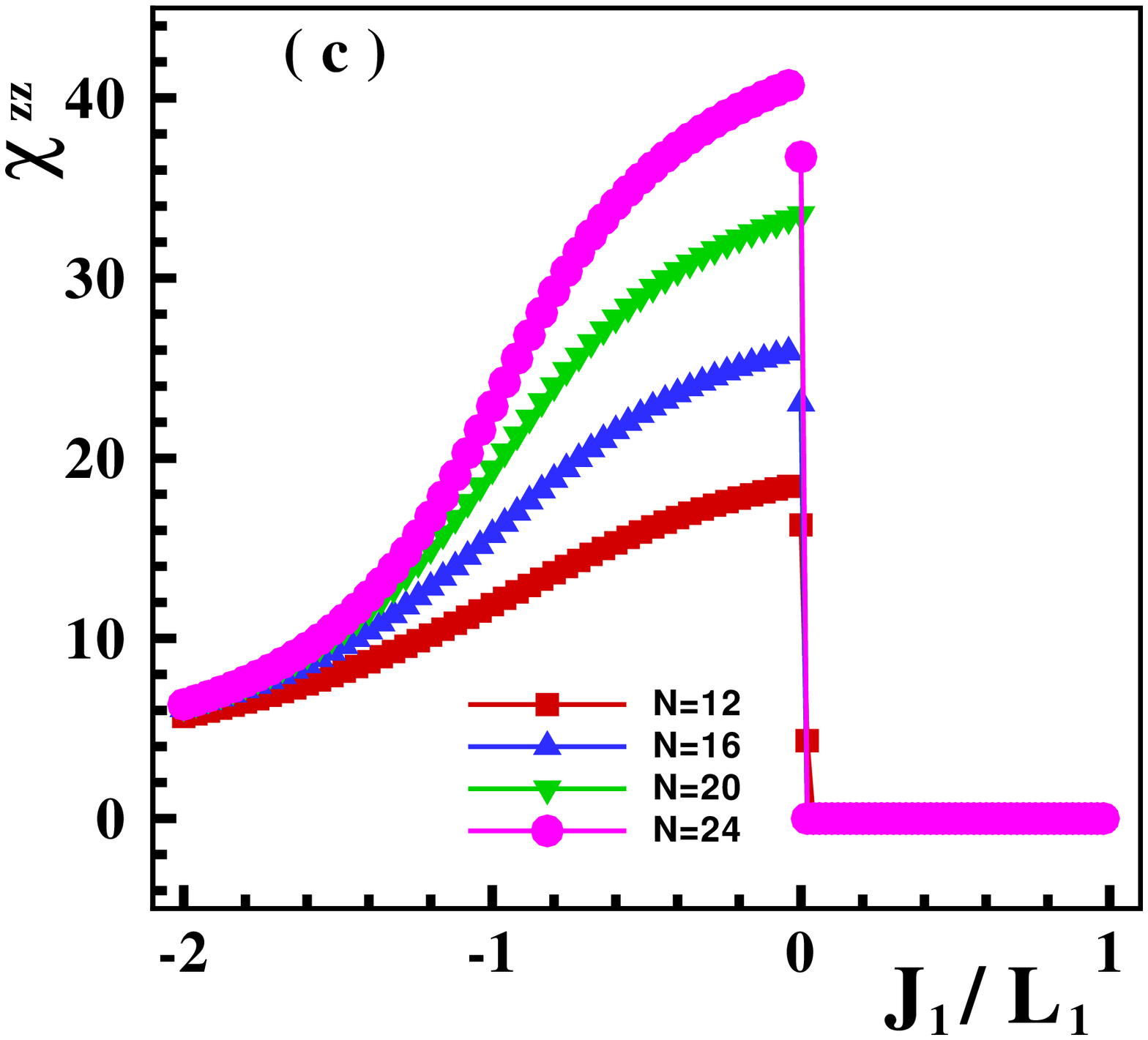}
} \caption{(Color online.) The correlation function of
stripe-antiferromagnetic order $\chi^{zz}$ as a function of $J_1$,
for chain lengths $N=12, 16, 20, 24$ and exchanges $L_{1}=1.0$,
(a)$J_2/L_1=2(1-J_1/L_1)$, (b)$J_{2}/L_1=1-J_1/L_1$ and (c)
$J_{2}/L_1=(1/2)(1-J_1/L_1)$ } \label{chi-zz}
\end{figure*}

It is known that the spin structure factor give us a deep insight
into the characteristics of the ground state\cite{Mori95}. In
Fig.\ref{correlations} we have plotted $W_{1}^{zz}$ as a function
of $n$ along the three paths. In the case of
$J_2/L_1=2(1-J_1/L_1)$ we expect three different magnetic phases
in the ground state phase diagram, indicated at the regions (I.):
$J_1/L_1<0$ , (II.): $0<J_1/L_1<0.5$ and (IV.): $0.5<J_1/L_1<1.0$.
In the case of $J_2/L_1=\frac{1}{2}(1-J_1/L_1)$, one additional
phase is expected in the region (III.): $-1.0<J_1/L_1<0$. The
selected values of the exchange $J_1/L_1$ in the figure cover all
regions of the ground state phase diagram. It can be seen from
this figure (Fig.\ref{correlations}) that the $z$ component of the
spins on odd sites are pointed in the same direction with the
$\sigma_{1}^{z}$ and others (on even sites) are pointed in
opposite direction  at $J_1/L_1=0.8$. This is an indication for
the N$\acute{e}$el ordering in the region (IV.) $J_1/L_1>0$,
$J_2/L_1<1$. It is surprising that the numerical results in
Fig.\ref{correlations}(c) at $J_1/L_1=-0.4$ suggest that a
different novel phase can be exist in the region (III.)
$J_1/L_1<0$, $J_2/L_1<1$. The values of the spin-spin correlation
function $W_{1}^{zz}$ in Fig.\ref{correlations}(c) at
$J_1/L_1=-0.4$, show that the two spins on the second odd bond are
pointed in the opposite direction with the $\sigma_{1}^{z}$ (or
similarly with the two spins on the first odd bond) but two spins
on the third odd bond are pointed in the same direction with the
$\sigma_{1}^{z}$. Clearly seen that the mentioned behavior
continue in whole of the chain system. We denote this novel kind
of the ordering in the ground state magnetic phase diagram of the
1D QCM by the "stripe-antiferromagnetic phase".

Classically, the effect of the negative exchange $J_1/L_1<0$ is
interesting. In the special case of $J_2/L_1=0$, the Hamiltonian
reduces to the alternating $ZZ$ Ising model. The ground state of
the alternating F-AF ZZ Ising model has a form $|\Psi_{Gs}
\rangle=|\uparrow\uparrow\downarrow\downarrow\uparrow\uparrow\downarrow\downarrow
...   \rangle$ with the long-range order canted spins in the
direction of the $z$ axis, as illustrated in Fig.\ref{Schematic}.
The ordering of this phase is a type of the
stripe-antiferromagnetic phase. Therefore, the order parameter of
the stripe-antiferromagnetic phase is defined
as\cite{Mahdavifar08}
\begin{eqnarray}
M_{sp}^{z} = \frac{2}{N}\langle
\sum_{j=1}^{N/2}(-1)^{j}(\sigma_{2j-1}^{z}+ \sigma_{2j}^{z})
\rangle.
\end{eqnarray}
For any value of the $J_1/L_1$, the Lanczos results lead to the
staggered magnetization,
$M_{st}^{z}=\frac{1}{N}\sum_{j=1}^{N}(-1)^{j}\sigma_{j}^{z}=0$,
and $M_{sp}^{z} =0$, since the ground state is degenerate and in a
finite system no symmetry breaking happens. However the spin-spin
correlation function diverge in the ordered phase as
$N\longrightarrow\infty$. We computed the spin structure factor
$S^{zz}(\pi)$ and the correlation function of the
stripe-antiferromagnetic order parameter given by
\begin{eqnarray}
\chi^{zz} = \langle \sum_{n=1}^{N/2-1}(-1)^{n}(\sigma_{2j-1}^{z}+
\sigma_{2j}^{z}) (\sigma_{2j-1+2n}^{z}+ \sigma_{2j+2n}^{z})
\rangle.
\end{eqnarray}

\begin{figure}[t]
\centerline{\psfig{file=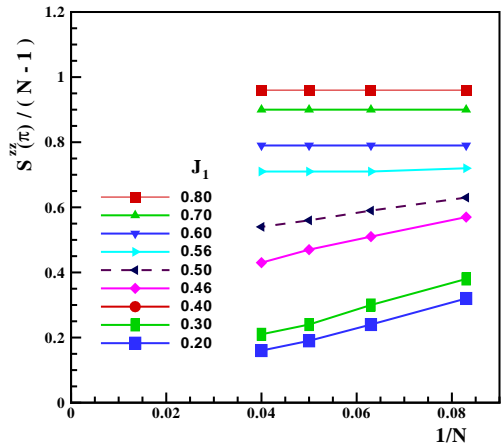,width=3.0in}} \caption{(Color
online.) The spin structure factor $S^{zz}(\pi)$ for $L_1=1.0$ and
$J_2/L_1=2(1-J_1/L_1)$ plotted as a function of the inverse chain
length $1/N$ for different values of the exchange $J_1$. The
dashed line which correspond to $J_1=0.5$ and $J_2=1.0$ marks the
second order quantum phase transition.} \label{N-Neel}
\end{figure}

\begin{figure}[t]
\centerline{\psfig{file=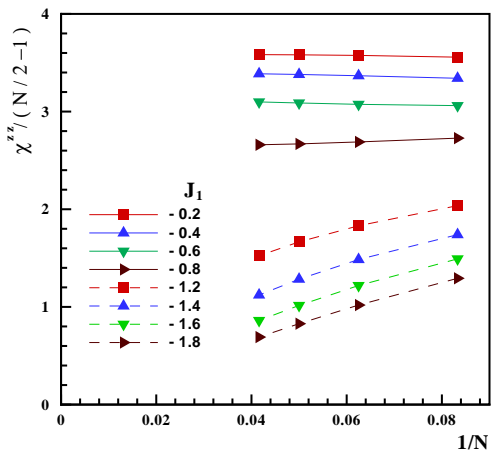,width=3.0in}}
\caption{(Color online.) The correlation function of the
stripe-antiferromagnetic order parameter $\chi^{zz}$ for $L_1=1.0$
and $J_2/L_1=1/2(1-J_1/L_1)$ plotted as a function of the inverse
chain length $1/N$ for different values of the exchange $J_1<0$.
The diminishing behavior of the numerical results (dashed lines)
in the region (I.): $J_1/L_1<-1$, $J_2/L_1>1$, shows that the
stripe-antiferromagnetic order in the region (I.) is not true
long-range (local ordering). } \label{N-Stripe}
\end{figure}

In Fig.\ref{Str-Neel}, we have plotted $S^{zz}(q=\pi)/(N-1)$ as a
function of $J_1/L_1$ for the chain lengths $N=12, 16, 20, 24$.
The $J_1/L_1$-dependency of the spin structure factor,
$S^{zz}(q=\pi)$, is qualitatively the same as the staggered
magnetization, $M_{st}^{z}$. It can be seen that in the region
(IV.) $J_1/L_1>0$, $J_2/L_1<1$, the ground state of the system is
in the N$\acute{e}$el phase. At $J_1/L_1=1.0$ (or $J_2/L_1=0)$,
the saturated N$\acute{e}$el phase is dominated in well agreement
with the $zz$ Ising model. Induced quantum fluctuations by
increasing $J_2/L_1$ from zero (or decreasing $J_1/L_1$ from the
value one), decreases the staggered magnetization from saturation
value. In all cases it is clearly seen that the staggered
magnetization as a function of the $J_1/L_1$, displays a jump to
zero at the critical value $J_1/L_1=0$ which is known as the
first-order phase transition. There is not the N$\acute{e}$el
ordering along $z$ axis in the region $J_1/L_1<0$. Overlapping of
the numerical results in all regions, except the region
$0<J_1/L_1\leq0.5$ and $J_2/L_1>1$ (Fig.\ref{Str-Neel}(a)), shows
a divergent behavior of the function $S^{zz}(\pi)$ by increasing
the size of chain $N$. This justifies that the N$\acute{e}$el
ordering along the $z$ axis is true long-range order in the region
(IV.) of the ground state phase diagram. To check the existence of
the N$\acute{e}$el order in the thermodynamic limit $N
\longrightarrow \infty$ of the system in the region (II.):
$0<J_1/L_1\leq0.5$, $J_2/L_1>1$ , we have plotted in
Fig.\ref{N-Neel} the $N$ dependence of $S^{zz}(\pi)/(N-1)$ for
different values of $J_1/L_1>0$. As is seen from this figure in
the region $0<J_1/L_1\leq0.5$, $J_2/L_1>1$ there is a diminishing
behavior which shows that the N$\acute{e}$el order in the region
(II.) is local (see also Fig.\ref{correlations}(a)). Therefore, we
conclude that the N$\acute{e}$el order, only exist in the region
(IV.) of the ground state phase diagram and a second-order quantum
phase transition happens at critical value $J_2/L_1=1.0$.
\begin{figure}
\centerline{\psfig{file=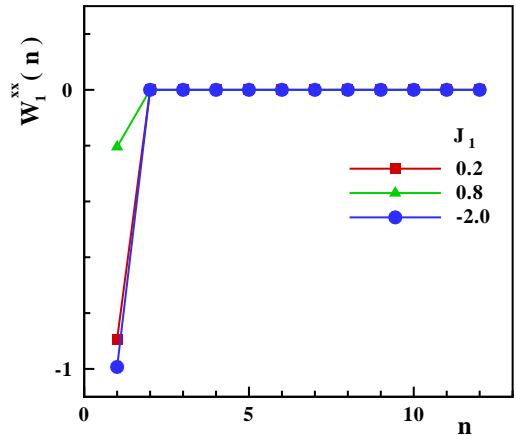,width=3.0in}}
\caption{(Color online.) The spin-spin correlation $W_{1}^{xx}(n)$
as a function of the $n$, for chain length $N=24$ and exchanges
$L_{1}=1.0$ and path $J_2/L_1=2(1-J_1/L_1)$.}
\label{XX-correlation}
\end{figure}

An additional insight into the nature of different phases can be
obtained by studying the correlation function $\chi^{zz}$. In
Fig.\ref{chi-zz}, we have plotted $\chi^{zz}$ as a function of the
exchange $J_1/L_1$ for different values of the chain length $N=12,
16, 20, 24$. In all cases it is clearly seen that there is no
stripe-antiferromagnetic ordering along the $z$ axis in the region
$J_1/L_1>0$ (regions II. and IV.). In contrast, in the region
(III.) $-1<J_1/L_1<0$, $J_2/L_1<1$ (see Fig.\ref{chi-zz}(c)), a
profound stripe-antiferromagnetic order exist in the $z$
direction. In the limit  $J_1/L_1\longrightarrow 0^-$ (or
$J_2/L_1\longrightarrow 0.5^+$), by checking the numerical value
of the $\chi^{zz}$ we found that the saturated
stripe-antiferromagnetic phase dominate in the ground state phase
diagram. The induced quantum fluctuations by increasing $J_2/L_1$
from $0.5$ (or decreasing $J_1/L_1$ from zero), decreases the
stripe-antiferromagnetic order from saturation value.
 Also, it is clearly seen that the stripe-antiferromagnetic order displays a jump to zero at the
 critical value $J_1/L_1=0$
 which is also an indication of  the first-order phase transition.
 To check the existence of the stripe-antiferromagnetic order in the
thermodynamic limit $N \longrightarrow \infty$ of the system in
the region (III.): $J_1/L_1<0$, $J_2/L_1<1$, we have plotted in
Fig.\ref{N-Stripe} the $N$ dependence of $\chi^{zz}/(N/2-1)$ for
different values of $J_1/L_1<0$. As is seen from this figure in
the region $-1.0<J_1/L_1<0$, $J_2/L_1<1$ there is a divergent
behavior which shows that the stripe-antiferromagnetic order in
the region (III.) is true long-range order. On the other hand the
diminishing behavior of the numerical results in the region (I.):
$J_1/L_1<-1$, $J_2/L_1>1$, shows that the stripe-antiferromagnetic
order in the region (I.) is local. Therefor, by investigating the
$N$ dependence of $\chi^{zz}/(N/2-1)$ for different values of
$J_1/L_1$, we found that there is no long-range
stripe-antiferromagnetic order in regions (I. and II. and IV.) and
only long-range stripe-antiferromagnetic order exist in the region
(III.). However, the numerical
 results on finite chains plotted in Fig.\ref{chi-zz}(b) , may suggest that there is the  stripe-antiferromagnetic
 long-range order in a very narrow
 region very close to the multicritical point $J_1/L_1=0$, $J_2/L_1=1.0$, and at this point jumps
 from a non-saturate value to zero.

To complete our numerical study of the ground state magnetic phase
diagram of the model, we have computed the $x$ component of the
spin-spin correlation function $W_1^{xx}$ in all cases. As an
example, in Fig.\ref{XX-correlation} we have plotted $W_{1}^{xx}$
as a function of $n$ along the path $J_2/L_1=2.0(1-J_1/L_1)$.
Clearly seen that along the $x$ axis, there are not any
conventional magnetic long-range order. We also found the same
results for other paths.

\section{conclusion}\label{sec-III }

In this paper we have studied the elementary excitation and the
magnetic ground state phase diagram of the 1D quantum compass
model. We have implemented the Lanczos method to numerically
diagonalize finite chains. Using the exact diagonalization
results, first we have calculated the energy gap and various
short-rang correlation functions
($\langle\sigma_{2i-1}^{x(z)}\sigma_{2i}^{x(z)}\rangle$ and
$\langle\sigma_{2i}^{x(z)}\sigma_{2i+1}^{x(z)}\rangle$). In
complete agreement with analytical results\cite{Eriksson09}, we
have showed that there are four regimes with different short-range
correlations, which are separated by lines of the first and
second-order phase transitions (see Fig.\ref{phase-diagram}). Then
to answers a very important question: "What are the long-range
ordered phases in the ground state phase diagram of the system?",
by plotting the spin-spin correlation functions, we have found
that two kind of magnetic long-range orders exist in the ground
state phase diagram, a type of the stripe-antiferromagnetic and
the N$\acute{e}$el orders in the regions (III.) $J_1/L_1<0$,
$J_2/L_1<1$ and (IV.) $J_1/L_1>0$, $J_2/L_1<1$, respectively. By a
detailed analysis of the numerical results on the spin structure
factors and the correlation function of the
stripe-antiferromagnetic order parameter, we have shown that the
N$\acute{e}$el and the stripe-antiferromagnet orderings are true
lang-range orders.

\section{acknowledgments}
Author would like to thank G. I. Japaridze, H. Johannesson, E.
Eriksson, H. Sadat Nabi and T. Vekua  for very useful comments and
interesting discussions.




\vspace{0.3cm}

\end{document}